\documentclass[usenatbib]{mn2e}
\usepackage{psfig}

\newcommand{\beq}{\begin{equation}}
\newcommand{\eeq}{\end{equation}}

\newcommand{\rem}[1]{{ }}

\def\apj{ApJ}
\def\mnras{MNRAS}

\def\aap{A\&A}

\bibpunct[,]{(}{)}{;}{a}{}{,}

\begin{document}

\title{Self-Similar Hot Accretion onto a Spinning Neutron Star: Matching the
Outer Boundary Conditions}
\author[R. Narayan and M.V. Medvedev]{
Ramesh Narayan$^{1}$ and 
Mikhail V. Medvedev$^{2}$ \\
$^1$ Harvard-Smithsonian Center for Astrophysics, 60 Garden Street, 
Cambridge, MA 02138\\
$^2$ Department of Physics and Astronomy, University of Kansas,
Lawrence, KS 66045
}

\maketitle

\begin{abstract}
Medvedev \& Narayan have described a hot accretion flow onto a
spinning neutron star in which the gas viscously brakes the spin of
the star.  Their self-similar solution has the surprising property
that the density, temperature and angular velocity of the gas at any
radius are completely independent of the outer boundary conditions.
Hence, the solution cannot be matched to a general external medium.
We resolve this paradoxical situation by showing that there is a
second self-similar solution which bridges the gap between the
original solution and the external medium. This new solution has an
extra degree of freedom which permits it to match general outer
boundary conditions.  We confirm the main features of the analytical
results with a full numerical solution.
\end{abstract}

\maketitle

\section{Introduction}

Accretion flows around compact objects frequently radiate significant
levels of hard X-rays.  This has provided strong motivation for the
study of hot accretion flows.  \citet{ZS69} and \citet{AW73}
considered spherically free-falling plasma impinging on the surface of
a neutron star (NS).  They calculated the penetration depth of the
falling protons and made preliminary estimates of the radiated
spectrum.  Their ideas were followed up by a number of later authors
(e.g., \citealp{Tur94,Zam95,Zan98}), who computed detailed spectra.
\citet{D01} modified the model of \citet{ZS69} by considering a
rotating advection-dominated accretion flow around a NS.  Their model
represents an improvement on the earlier work since it includes
angular momentum and viscosity.

A fluid approach to hot accretion onto a NS was pioneered by
\citet{SS75}, who worked out the structure of the standing shock in a
spherical flow, and computed the two-temperature structure and the
resulting radiation spectrum of the post-shock gas.  The equivalent
problem for an accreting white dwarf was analyzed by \citet{KL82}.  In
related work, \citet{CS97} described the hydrodynamics of spherical
accretion onto black holes and NSs, but without including radiation
processes. 

The above studies involve flows in which the accreting matter crashes
on the surface of the star, forming a discontinuity or a shock of some
kind.  Recently, \citet[ hereafter MN01]{MN01} discovered a rotating
solution of the viscous fluid equations that corresponds to hot
quasi-spherical accretion onto a spinning NS.  Their solution is closely
related to the two-temperature flow described by \citet{SLE76}.  A
feature of the MN01 solution is that the gas moves subsonically in the
radial direction and merges with the accreting star without a shock.
The flow essentially ``settles'' onto the rotating star; the solution
may thus be referred to as a ``hot settling flow'' (hot because the
gas is at the virial temperature and has a quasi-spherical
morphology).  MN01 showed that the accreting gas removes angular
momentum from the central star and that this braking action dominates
the energy equation of the accreting gas.  The flow could thus also be
called a ``hot brake.''  The hot settling flow should not be confused
with the boundary layer which forms close the stellar surface (e.g.,
\citealp{NP93}), where the gas density is high and steep spatial
gradients are present. The hot settling flow forms {\it outside} the
boundary layer and extends radially to a large distance, typically
thousands of stellar radii or more (see MN01).  \citet{I01,I03},
following up on earlier work by \citet{DP81}, has recently described a
subsonic hot accretion flow around a magnetized neutron star in the
propeller state.  The relation between his solution and our hot
settling flow is discussed in \S4.

The relevance of the hot settling flow to real systems is presently
unclear, though several accreting white dwarf and black hole systems
have been suggested as candidates for such a rotation-powered flow
\citep{MMen02,MMur02}. While MN01 described the properties of the
self-similar region of the flow, they did not discuss how to match the
solution to realistic boundary conditions.  The matching to an
external medium at large radii is particularly problematic, since the
self-similar settling flow solution has the remarkable property that
the density, temperature and angular velocity of the gas at given
radius depend only on the dimensionless spin of the central star, and
are completely independent of the properties of the external medium.
This leads to an apparently serious problem.  If one tries to match
the solution to the external medium by selecting the radius at which
the pressure of the solution matches that of the medium, then neither
the density nor the temperature will agree.  Does this mean that the
solution is physically inconsistent?  We argue otherwise in this
paper.

We show that there exists a second self-similar solution at radii
outside the MN01 hot settling flow, which acts as a bridge between the
MN01 solution and the external medium.  This bridging solution has an
extra degree of freedom which allows it to match a general density and
temperature in the external medium.  Apart from the additional degree
of freedom, the solution retains many of the features of the MN01
settling flow: (i) it is pressure supported, (ii) it resembles a
static atmosphere at low mass accretion rates, and (iii) the angular
momentum flux associated with the viscous braking of the central star
dominates most of the physics.

The paper is organized as follows. In \S\ref{s:th} we derive
analytical expressions for various self-similar solutions.  In
\S\ref{s:num} we present full numerical solutions and compare them
with the analytical solutions.  Finally, in \S4 we conclude with a
brief discussion.

\section{Analytical Self-Similar Solutions}
\label{s:th}

We consider gas accreting viscously onto a compact spinning object.
The central object has a radius $R_*$, a mass
$M_*=m M_{\sun}$, and an angular velocity $\Omega_*=s \Omega_K(R_*)$,
where $\Omega_K(R)=(GM/R^3)^{1/2}$ is the Keplerian angular velocity
at radius $R$.  We measure the accretion rate in Eddington units,
$\dot m=\dot M/\dot M_{\rm Edd}$, and the radius in Schwarzchild
units, $r=R/R_g$, where $\dot M_{\rm
Edd}=1.39\times10^{18}m$~g~s$^{-1}$ (corresponding to a radiative
efficiency of 10\%) and $R_g=2GM/c^2$.

We assume that the flow is hot and quasi-spherical, which generally
requires a low mass accretion rate (see Narayan et al. 1997). The
accreting gas has nearly the virial temperature, i.e., $c_s^2\sim GM/R
\sim (\Omega_K R)^2$, and the local vertical scale height
$H=c_s/\Omega_K$ is comparable to the local radius $R$.  We may then
use the height integrated hydrodynamic equations for a steady,
rotating, axisymmetric flow, and for simplicity we may set $H= R$ (see
MN01).  (Note that, even when $H\approx R$, the stability properties
of the flow may depend on whether we use $H$ or $R$ in the equations;
this is briefly discussed in \S4).  In the following, we closely
follow the analysis of MN01, with a few changes.

For simplicity, we set $\dot m=0$ and omit the continuity equation.
Thus, the gas configuration corresponds to a radially static
``atmosphere.''  The motivation for this approximation follows from
the observation that the density $\rho$, temperature $T$ and the
angular velocity $\Omega$ of the gas in the MN01 self-similar solution
are completely independent of $\dot m$.  Only the radial velocity $v$
depends on $\dot m$, and it is given trivially by the spherical
continuity equation \beq v={\dot M\over 4\pi R^2\rho}.
\label{v}
\eeq Because of this, we do not lose any generality by setting $\dot
m=v=0$ in the analysis; we may always introduce a finite $\dot m$ and
finite $v$ after the fact.

We assume that the flow is highly sub-Keplerian, $\Omega(R)
\ll\Omega_K(R)$, so that the centrifugal support is negligible
compared to the pressure support.  The radial momentum equation then
takes the following simple form, \beq {GM\over R^2}=-{1\over
\rho}{d(\rho c_s^2)\over dR},
\label{mmtm}
\eeq where we have used the fact that $v\sim0$ and written the
pressure as $p=\rho c_s^2$ where $c_s$ is the isothermal sound speed.
MN01 present a more complete analysis in which they do not assume that
the rotation is slow.  They then obtain an extra factor of $(1-s^2)$
in their equation, which propagates through to all the results.  Since
we ignore the factor, our analysis corresponds to the case of a
slowly-spinning star: $s^2\ll1$.  This approximation is made only to
simplify the analysis, and all the results may be generalized for
arbitrary $s$.

From the analysis in MN01, we know that the accreting gas in our
problem acts as a brake on the central spinning star and transports
angular momentum outward through the action of viscosity. We therefore
write the angular momentum conservation equation for the gas as
follows, \beq \dot J=4\pi\nu\rho R^4{d\Omega\over dR}={\rm constant},
\label{jdot}
\eeq where $\dot J$ is the outward angular momentum flux, and $\nu$ is
the kinematic coefficient of viscosity. This equation is exactly valid
in steady state if $\dot m=0$. When $\dot m$ is non-zero, there is an
additional term, $\dot M\Omega R^2$, due to the flux of angular
momentum carried in by the accreting gas.  The key feature of the MN01
hot brake solution is that the latter flux is negligible compared to
the outward flux from the star.  Equation (\ref{jdot}) is, therefore,
valid even when $\dot m\ne0$, so long as $\dot m$ is small enough for
the term $\dot M \Omega R^2$ to be negligible.

We employ the usual $\alpha$ prescription for the kinematic
coefficient of viscosity, which we write as \beq \nu=\alpha c_s H
\approx \alpha c_s R.
\label{nu}
\eeq Often, in accretion problems, one makes use of the relation
$H=c_s/\Omega_K$ and writes $\nu=\alpha c_s^2/\Omega_K$.  This
prescription is equivalent to equation (4) in the regime of the MN01
hot settling flow.  However, in the outer regions of the flow, where
the two new solutions described in \S\S2.2,2.3 apear, $H$ is much less
than $c_s/\Omega_K$, and $\nu=\alpha c_s^2/\Omega_K$ is not a good
approximation.  Equation (4) is a superior prescription and is
physically better motivated over a wide range of conditions (so long
as the flow is quasi-spherical).

For simplicity, we assume that the gas is one-temperature; it is
straightforward to generalize to the two-temperature regime, as done
in MN01, but the one-temperature analysis suffices for the present
paper. Hence do not need to treat electrons and protons separately.

Viscous braking heats the accreting gas, and we assume a steady state
in which this heating is balanced by cooling.  The energy conservation
equation thus becomes \beq q^+=q^-,
\label{energy}
\eeq where the viscous heating rate per unit volume $q^+$ and the
radiative cooling per unit volume via bremsstrahlung $q^-$ are given
by (MN01)
\begin{eqnarray*}
q^+&=&\nu\rho R^2\left( {d\Omega\over dR}\right)^2, \\
q^-&=&Q_{\rm ff,NR}\,\rho^2 \left({kT\over m_ec^2}\right)^{1/2}, \qquad
Q_{\rm ff,NR} = 5\sqrt{2}\pi^{-3/2}\alpha_f \sigma_T
{m_e c^3 \over m_p^2},
\end{eqnarray*}
where $\alpha_f$ is the fine structure constant, $\sigma_T$ is the
Thomson cross-section, and $m_e$ and $m_p$ are the mass of the
electron and the proton.  Equation (\ref{energy}) is exactly valid for
a steady state flow with $\dot m= v=0$.  When $v \ne 0$, the energy
equation has another term corresponding to the advection of energy.
In advection-dominated accretion flows, for instance, this term
dominates over the cooling term $q^-$ \citep{NMQ97}.  In the present
case, however, we consider a situation in which the advection term is
negligible (which corresponds to low $\dot m$).

Finally, we assume that the spinning star is immersed in a uniform
external medium with a density $\rho_{\rm ext}$, temperature $T_{\rm
ext}$ and pressure $p_{\rm ext}$.  We seek an accretion flow solution
that extends from the spinning star on the inside to the external
medium on the outside.  As we show below, the solution consists of two
distinct self-similar regimes, plus a third asymptotic regime inside
the external medium.

\subsection{Inner self-similar solution}
\label{s:1st}

We first consider the inner regions of the flow, where the pressure
$p\gg p_{\rm ext}$.  This is the regime of the MN01 solution, where
the variables have the following radial dependences: \beq
\rho=\rho_1r^{-2}, \quad T=T_1r^{-1}, \quad \Omega=\Omega_1r^{-3/2}.
\label{1st-sol}
\eeq The subscript ``1'' in the coefficients is to indicate that this
is the first solution, to distinguish it from the second and third
solutions described below.  By substituting the above solution in
equations (\ref{mmtm}), (\ref{jdot}) and (\ref{energy}), we see that
it satisfies the basic conservation laws.  We may also solve for the
numerical constants:
\begin{eqnarray*}
\rho_1&=&\frac{\alpha s^2}{R_g}\,\frac{9}{2^{5/2}}
\left({\frac{m_e}{m_p}}\right)^{1/2}c^3Q_{\rm ff,NR},\\ kT_1&=&{m_p
c^2 \over 12},\\ \Omega_1&=&{s\,c \over \sqrt{2}R_g}=s\,\Omega_K(R_g).
\end{eqnarray*}
We note that if $\dot m \ne 0$ then the flow has a small
constant radial velocity: \beq v\propto r^0, \eeq as follows from
equation (\ref{v}).

The angular momentum flux in the solution is given by \beq \dot
J=-\alpha^2 s^3 R_g^2\,\frac{3^{5/2}}{2^{5/2}}
\left({\frac{m_e}{m_p}}\right)^{1/2}\frac{c^5}{Q_{\rm ff,NR}}.
\label{jdot1}
\eeq By assumption, this flux is much greater than the angular
momentum flux due to accretion, which sets an upper limit on the mass
accretion rate for the solution to be valid (see MN01).  The pressure
is given by \beq p =\rho c_s^2=\rho_1 c_{s1}^2\,r^{-3} \equiv
p_1\,r^{-3},
\label{1st-true}
\eeq where $c_{s1}^2=2kT_1/m_p$.

The above self-similar solution describes the flow at radii $r\ll
(p_1/p_{\rm ext})^{1/3}$, where the pressure $p \gg p_{\rm ext}$.  As
mentioned in \S1, the solution has the remarkable property that all
the quantities are uniquely determined by a single parameter $s$ ---
the dimensionless spin of the central object --- specified on the
inner boundary.  The fact that the solution does not depend on the
outer boundary condition in any way means that there is no simple way
to match it to the external medium.  Clearly, there has to be a second
solution to bridge the gap between this solution and the external
medium.  We derive the bridging solution in the next subsection.

\subsection{A second self-similar solution}
\label{s:2nd}

We consider next the gas that lies just outside the region of validity
of the first self-similar solution described above.  In this zone, the
pressure is expected to be approximately equal to the external
pressure $p_{\rm ext}$: \beq \rho c_s^2 = p_{\rm ext} = {\rm
constant}.
\label{mmtm2}
\eeq This condition replaces the hydrostatic equilibrium equation
(\ref{mmtm}), while equations (\ref{jdot}) and (\ref{energy}) continue
to be valid.  In this region, we find that there is a second
self-similar solution of the form \beq \rho=\rho_2r^{-7/2}, \quad
T=T_2r^{7/2}, \quad \Omega=\Omega_2r^{-9/4},
\label{2nd-sol}
\eeq where the label ``2'' refers to the fact that this is our second
solution.

To match the second and first solutions, we require that the fluxes of
angular momentum in the two solutions must be equal; this yields the
constraint
$(3/2)\rho_1\Omega_1T_1^{1/2}=(9/4)\rho_2\Omega_2T_2^{1/2}$.  Making
use of this and the other equations, we solve for the numerical
coefficients in equation (\ref{2nd-sol}):
\begin{eqnarray*}
\rho_2&=&\frac{\alpha^{3/2} s^3}{p_{\rm ext}^{1/2}R_g^{3/2}}\,
\frac{3^{5/2}}{2^{9/4}}\left(\frac{m_e}{m_p}\right)^{3/4}
\frac{c^4}{Q_{\rm ff,NR}^{3/2}},\\
kT_2&=&\frac{p_{\rm ext}^{3/2}R_g^{3/2}}{\alpha^{3/2} s^3}\,
\frac{2^{5/4}}{3^{5/2}}\left(\frac{m_p}{m_e}\right)^{3/4}
\frac{m_p Q_{\rm ff,NR}^{3/2}}{c^4},\\
\Omega_2&=&\frac{\alpha^{1/4} s^{3/2}}{p_{\rm ext}^{1/4} R_g^{5/4}}\,
2^{-3/8}3^{-3/4}\left(\frac{m_e}{m_p}\right)^{1/8}
\frac{c^2}{Q_{\rm ff,NR}^{1/4}}.
\end{eqnarray*}
The pressure in this solution is constant and equal to the external
pressure, $p_{\rm ext}$, and the angular momentum flux is also
constant and is equal to $\dot J$ in equation (\ref{jdot1}).  If the
flow has a small but nonzero accretion rate, $\dot m\not=0$, then its
radial velocity varies as [see eq. (\ref{v})] \beq v\propto r^{3/2}.
\eeq

Whereas the original MN01 self-similar solution has a unique profile
for a given choice of $s$, we see that the second solution derived
here has an extra degree of freedom, namely the external pressure
$p_{\rm ext}$. This extra degree of freedom solves the problem
discussed in \S1.  Thus, the full solution consists of two zones: an
inner zone described by the first (MN01) solution (\ref{1st-sol}) and
an outer zone described by the second solution (\ref{2nd-sol}).  The
radius $r_{\rm match}$ at which the two solutions match is obtained by
equating the pressures: \beq r_{\rm match}=\frac{\alpha^{1/3}
s^{2/3}}{p_{\rm ext}^{1/3} R_g^{1/3}}\, \frac{3^{1/3}}{2^{7/6}}
\left({\frac{m_e}{m_p}}\right)^{1/6}c^{5/3}Q_{\rm ff,NR}^{1/3}.
\label{R12}
\eeq

The second solution matches the external medium at the radius $r_{\rm
ext}$ at which its temperature matches that of the medium.  This gives
\beq r_{\rm ext}= \frac{\alpha^{3/7} s^{6/7}}{p_{\rm
ext}^{3/7}(kT_{\rm ext})^{2/7}R_g^{3/7}}\,
\frac{3^{5/7}}{2^{5/14}}\left(\frac{m_e}{m_p}\right)^{3/14}
\frac{c^{8/7}}{m_p Q_{\rm ff,NR}^{3/7}}.
\label{R23}
\eeq If we wish we could also write this in terms of the external
density by making the substitution $kT_{\rm ext} = m_p p_{\rm ext}
/2\rho_{\rm ext}$.

\subsection{Asymptotic solution in the external medium}
\label{s:3rd}

For completeness, we present here the solution inside the external
medium.  By assumption, the external medium has a uniform temperature
and density, and a uniform rate of cooling.  To maintain equilibrium,
there has to be some constant source of heat that exactly compensates
for the cooling.  We assume that such a source of heat exists (e.g.,
cosmic rays).  The rotation $\Omega$ is non-zero, but it decays
rapidly outward.  The small amount of rotation helps to transport the
angular momentum flux from the star out into the external medium.
Solving the angular momentum conservation law (3), we obtain the
following solution \beq \rho=\rho_{\rm ext}, \quad T=T_{\rm ext},
\quad \Omega=\Omega_3 r^{-4},
\label{3rd-sol}
\eeq where \beq \Omega_3 = \alpha\, s^3\,\frac{3^{5/2}}{2^7\pi}
\frac{m_e^{1/2} c^5}{R_g^2 Q_{\rm ff,NR}}\, \rho_{\rm ext}^{-1}\left(k
T_{\rm ext}\right)^{-1/2} , \eeq and $p_{\rm ext}=2kT_{\rm
ext}\rho_{\rm ext}/m_p$.  For $\dot m\not=0$, the velocity scales as
$r^{-2}$.

\section{Numerical Results}
\label{s:num}

The three self-similar solutions written above are special solutions
of the basic differential equations (2), (3) and (5), which are valid
under specific conditions.  To check the validity of these analytical
solutions, we have computed numerical solutions of the basic
differential equations.  We use the same code as in MN01, with two
changes.  First, we switched to the viscosity prescription given in
equation (\ref{nu}), rather than the prescription $\nu = \alpha
c_s^2/\Omega_K$ used in MN01.  Second, in addition to viscous heating,
we included a constant heating rate which we adjusted so as to balance
the radiative cooling in the homogeneous external medium (see \S2.3).
The code uses a relaxation method to solve the one-dimensional
hydrodynamic equations with specified inner and outer boundary
conditions.  Although it employs the full equations of a
two-temperature plasma, the results are essentially equivalent to
those of a one-temperature plasma in the region of interest for this
paper, namely the region at large radius where the flow matches onto
the external medium

\begin{figure}
\psfig{file=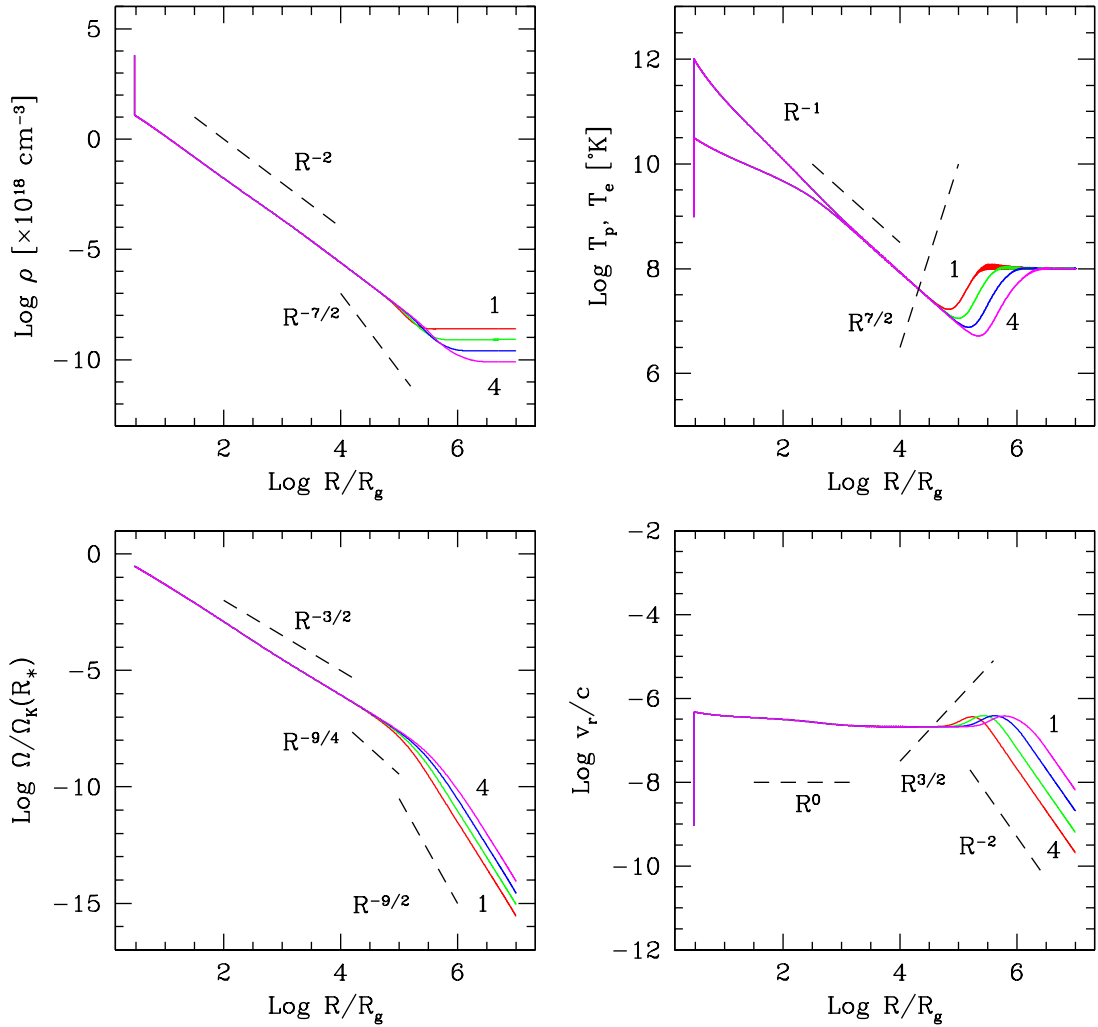,width=6in}
\caption{Profiles of density (top left panel), temperature (top right
panel, the electron temperature is the lower curve on the left and the
proton temperature is the higher curve), angular velocity (bottom left
panel), and radial velocity (bottom right panel), for four numerical
solutions of the full height-integrated differential equations.  The
four solutions correspond to different values of the density of the
external medium: $\rho_{\rm ext}= 2.5\times10^9,\ 8.1\times10^8,\
2.5\times10^8,\ 8.1\times10^7$~cm$^{-3}$.  The first and fourth
solutions are labeled 1 and 4, respectively.  The temperature of the
external medium and the accretion rate are kept fixed in all the
solutions: $T_{p,{\rm ext}}=T_{e,{\rm ext}}=10^8$K, $\dot m=2\times
10^{-5}$.  The analytical slopes of the three self-similar solutions
described in \S\S\ref{s:1st}--\ref{s:3rd} are shown for comparison
\label{f:prof}}
\end{figure}

In the calculations, the flow was taken to extend from an inner radius
$R_{\rm in}=3~R_g$ to $R_{\rm out}=10^7~R_g$.  The mass accretion rate
was taken to be low, $\dot m=2\times 10^{-5}$, in order that the flow
should correspond to the regime of the hot settling flow solution.  We
took the viscosity parameter to be $\alpha=0.1$ and set the spin of
the star to be $s=0.3$ (i.e., 30\% of the Keplerian rotation at the
stellar surface).  We took the other inner boundary conditions to be
the same as in MN01.  At the outer boundary, we specified the
temperature and density of the external medium.  Figure \ref{f:prof}
shows four solutions.  The external temperature is kept fixed at
$T(R_{\rm ext})=10^8$~K in all the solutions, but the external density
varies by a decade and a half: $\rho(R_{\rm ext})= 2.5\times10^9,\
8.1\times10^8,\ 2.5\times10^8,\ 8.1\times10^7$~cm$^{-3}$.  We have
also done other calculations in which we kept $\rho_{\rm ext}$ fixed
and varied $T_{\rm ext}$.  These give very similar results.

Fig. \ref{f:prof} shows that, right next to the star, there is a
boundary layer, where the density rises sharply as one goes into the
star and the temperature drops suddenly.  We do not analyze this
region.  Once we are outside the boundary layer, the gas behaves very
much according to the analytical solutions discussed in \S2.  Starting
just outside the boundary layer and extending over a wide range of
radius, the numerical solution exhibits a self-similar behavior with
power-law dependences of the density, temperature and angular
velocity.  This region corresponds to the self-similar solution of
MN01.  There are, in fact, two zones, an inner two-temperature zone,
and an outer one-temperature zone (MN01).  The latter corresponds to
solution 1 (eq. \ref{1st-sol}) discussed in \S2.1. The most notable
feature of this region is that the density, temperature and angular
velocity of the numerical solutions are completely independent of the
outer temperature and density, as predicted by the analytical
solution.  The slopes of the numerical curves also agree well with the
analytical scalings.

At a radius $R_{\rm match}\sim5\times10^4 ... 2\times10^5~R_g$
9depending on the outer pressure, see eq. \ref{R12}), solution 1
merges with solution 2 (eq. \ref{2nd-sol}) described in \S2.2.  Here,
the solution does depend on the outer boundary conditions, and it
scales roughly according to the slopes derived analytically.  At even
larger radii $R > R_{\rm ext}\sim3\times10^5 ... 2\times10^6~R_g$ (see
eq. \ref{R23}), the flow matches onto the ambient external medium.  In
this region we have solution 3 (eq. \ref{3rd-sol}) described in \S2.3.
As expected, out here only the angular velocity and the radial
velocity vary with radius.  Both have the scalings predicted for
solution 3.

\section{Discussion}
\label{s:concl}

In this paper, we have removed one piece of mystery surrounding the
self-similar ``hot settling flow'' or ``hot brake'' solution
discovered by MN01.  Specifically, we have shown that the remarkable
insensitivity of the MN01 solution to external boundary conditions is
a consequence of the fact that the solution is insulated from the
outer boundary by the presence of a second solution, which bridges the
gap between the first solution and the external medium.  We derived
the form of the second solution analytically in \S2.2 and showed via
numerical computations (\S3, Fig. 1) that the two solutions together
are able to match a wide range of outer boundary conditions.  This
solves one of the mysteries associated with the hot settling flow
solution.

There are, however, two other problems that still need to be
addressed.  First, the solution we have derived treats the mass
accretion rate $\dot m$ as a free parameter.  (Indeed, the analytical
solutions were obtained for the limit $\dot m\to0$, i.e. for a hot
atmosphere.)  What determines $\dot m$?  It is certainly not the outer
boundary, since we have obtained the complete outer solution.  The
accretion rate must therefore be determined by an inner boundary
condition.  This is not unexpected.  In the case of spherical
accretion, one recalls that, while the accretion rate for the
transonic solution is determined by the outer boundary conditions, the
accretion rate for the subsonic settling solution is determined by an
inner boundary condition.  In that problem, a whole family of settling
solutions exists.  Each member of the family has a different mass
accretion rate, and it is the manner in which the gas cools and
condenses on the accreting star that determines which particular
solution, i.e., which $\dot m$, is selected.  We expect the same
situation to apply to our problem.  Unfortunately, this means that in
order to estimate $\dot m$ we have to solve the full coupled
hydrodynamic and radiative transfer equations in the boundary layer
region next to the neutron star, with proper boundary conditions.  We
have not yet succeeded in this difficult exercise.

The second problem that needs to be addressed is the stability of the
hot settling flow solution.  Because the solution is hot, optically
thin and satisfies the energy balance condition (\ref{energy}), it is
similar in many respects to the hot solution discovered by Shapiro et
al. (1976).  The latter is known to be thermally very unstable (e.g.,
\citealp{P78}), and so one wonders whether the hot settling flow might
also be unstable. The thermal stability depends, in particular, on the
viscosity prescription chosen, as illustrated below (we thank the
referee for this simple argument).  With the prescription $\nu=\alpha
c_s^2/\Omega_K$ the flow is unstable because $q^+\propto T$ while
$q^-\propto\sqrt T$. Hence, a local increase of temperature results in
an increase of the net heating rate $q^+ - q^-$ which leads to a
further temperature increase.  However, for the viscosity prescription
given in equation (\ref{nu}), we have $q^+\propto\sqrt T\propto q^-$,
so that the flow is marginally stable. A more detailed analysis of the
problem is beyond the scope of the present paper.

We finally comment on the relationship of the hot settling flow
discussed in this paper to the subsonic propeller flow described by
\citet{I01,I03}.  Both flows describe the braking action of hot gas on
a spinning neutron star.  The propeller flow has been discussed in
connection with a strongly magnetized neutron star, while the hot
settling flow was developed to model accretion onto an unmagnetized
neutron star, but this is not a large distinction.  The main
difference between the two solutions is in the treatment of the energy
equation.  In Ikhsanov's subsonic propeller flow, the heating rate of
the accreting gas through viscous dissipation is much larger than the
radiative cooling rate.  The gas becomes convective and isentropic,
with density falling as $r^{-3/2}$.  The solution is in some sense
related to an advection-dominated accretion flow (Narayan et al. 1997)
or a convection-dominated accretion flow (Narayan, Igumenshchev \&
Abramowicz 2000; Quataert \& Gruzinov 2000).  In contrast, the hot
settling flow, as well as the two other solutions described in this
paper, satisfy detailed energy balance at each radius.  The viscous
heating at each point is exactly balanced by local radiative cooling
via optically thin bremsstrahlung, as indicated in equation (5) of the
present paper.  Furthermore, the density falls off as $r^{-2}$ and
$r^{-7/2}$ for solutions 1 and 2 rather than $r^{-3/2}$, the entropy
of the gas increases outward, and the gas is convectively stable
(MN01).

\bigskip
The authors are grateful to the referee for helpful suggestions, and
RN thanks Lars Hernquist for useful discussions.  This work was
supported in part by NASA grant NAG5-10780.

\onecolumn

\end{document}